# Reconfigurable qubit states and quantum trajectories in a synthetic artificial neuron network with a process to direct information generation from co-integrated burst-mode spiking under non-Markovianity

**Authors:** Osama M. Nayfeh[1†], Chris S. Horne[1]

**Affiliations:** [1] Naval Information Warfare Center (NIWC) Pacific, San Diego, CA USA

**†Corresponding Author:** osama.m.nayfeh.civ@us.navy.mil

## Abstract:

A synthetic artificial neuron network functional in a regime where quantum information processes are co-integrated with spiking computation provides significant improvement in the capabilities of neuromorphic systems in performing artificial intelligence and autonomy tasks. This provides the ability to execute with the qubit coherence states and entanglement as well as in tandem to perform functions such as read out and basic arithmetic with conventional spike-encoding. Ultimately, this enables the generation and computational processing of information packets with advanced capabilities and an increased level of security in their routing. We now use the dynamical pulse sequences generated by a memristive spiking neuron to drive synthetic neurons with built-in superconductor-ionic memories built in a lateral layout with integrated Niobium metal electrodes as well as a gate terminal and an atomic layer deposited ionic barrier. The memories operate at very low voltage and with direct, and hysteretic Josephson tunneling and provide enhanced coherent properties enabling qubit behavior. We operated now specifically in burst mode to drive its built-in reconfigurable qubit states and direct the resulting quantum trajectory. We analyze the new system with a Hamiltonian that considers an integrated rotational dependence, that's dependent on the unique co-integrated bursting mode spiking - and where the total above threshold spike-count is adjustable with variation of the level of coupling between the neurons. We then examined the impact of key parameters with a longer-term non-Markovian quantum memory and finally explored a process and algorithm for the generation of information packets with a coupled and entangled set of these artificial neuron qubits that provides for a quantum process to define the level of regularity or awareness of the information packets. These results therefore enable quantum neural networks where qubit/quantum memory states and the associated quantum trajectories are now available for conducting advanced computational algorithms in conjunction with the information processing capabilities of general spiking neurons.

## Introduction and motivation for biological-inspired synthetic artificial neuron network

A biologically inspired design of an artificial neuron network that can function with co-integrated spiking modes and coherent qubit states enables information processing that's not possible with conventional neuromorphic and quantum computing architectures. There is also emerging evidence in the literature that biological brains may operate with information processing mechanisms that combine traditional spiking neural computations that are co-integrated with quantum qubit operations for performing advanced information processing functionality *(1,2)*. There's also a concurrent multidisciplinary goal to help provide a clearer understanding of all the mechanisms for how our biological brains process information to help us better design artificial and synthetic systems that can help us in our everyday tasking as well as the discovery of technologies that can more properly interface and enhance our capabilities *(3)*. The neural architecture of the biological brains of humans as well as various animals spanning ground squirrels, lions, as well as bears, wolves and birds etc. provide an optimal starting point that is helpful in us designing technology as it has proven effective in both information processing as well as in its inherent secure information generation capabilities *(4,5)*. In this manuscript, we now take further steps than done previously in *(2)* and focus on observing and reconstructing the reconfigurable qubit-state dynamics when our quantum neural networks are excited with burst-mode spiking sequences that result in quantum trajectories that provide capability for a spiking neuromorphic circuit to also implement quantum information algorithms and enable the use of qubit entanglement and the emerging concepts of non-Markovianity in novel applications *(6-10)*.

Fig. 1(**A**) shows a modern concept diagram describing the biological architecture and the energy-level description Fig. 1(**B**) of the basic connections of a set of biological neurons with annotations to the design structure to describe how qubit-state processing is expected to occur in conjunction with the processes for traditional spiking neuron understanding. This includes the presence of electronics and chemical synaptic junctions *(11)*, typically a few nanometers in dimensions between them and that incorporate ionic channels containing i.e. sodium/potassium/calcium that mediate and drive the inter-connections between neurons as well as dendritic integration *(12)* with integrated microtubules containing protein electrostatic dimer molecules that further facilitates feedback processes. Such conceptual diagrams are a hot topic in the literature for approximating the spiking responses as well as exploring how biological ion channels and other elements such as foundational microtubules with special protein dimers may provide or qubit-state dynamics *(13)*. While many unknowns remain, it is useful for making this comparison to understand the response of biological neurons as well as fulfill the design goals for an artificial synthetic neuron that is operable in both the classical and quantum regimes. Whether or not quantum processes participate in biological brain activity is still an open question and especially to what degree if so, i.e. while quantum tunneling *(14)* effects seem very likely due to experimental observations- whether the implementation of quantum algorithms and processes such as coherent states and entanglement participate or not are still not fully determined, but may provide plausible explanations to the capabilities we and other species have in information processing tasks and the efficiency. Regardless, such postulations provide for an interesting concept for artificial and synthetic systems that may enable high performance information technologies not possible with their emulation strictly on digital machines or by stand-alone quantum computers and require such brain-inspired configuration for optimal design and provides also a route for miniaturization into practical configurations as today's quantum computers and datacenters require large areas and extensive energy resources *(15)*.

## Reconfigurable qubit-states/quantum trajectories with co-integrated burst-mode spiking

We examine a special case of our quantum artificial neural network hardware in a architecture where spiking neuron network hardware with superconductor-ionic memory devices in a sandwich structure such as described in *(2,16)* are used with adaptive-itinerant spiking in burst mode is used as the driving/excitation feature for built-in qubits designed with a nanofabricated qubit/quantum memory structure in a lateral format with a third terminal gating mechanism that can accept the burst-mode spiking as the coupling/excitation source . A circuit schematic is shown in Fig. 1(**C**) as well as a scanning electron micrograph of the nanofabricated structure. An energy-band diagram that describes the operation of the quasiparticles and cooper pairs in such a structure is shown in Fig 1(**D**) and a conductance measurement that demonstrates the sharp quantum tunneling behavior with hysteresis is shown in Fig. 1(**E-G**). To calibrate the burst-mode spiking used as the excitation, we optimize the circuit parameters/biasing and coupling strength to enter a burst-spiking mode Fig. 2 (**A-B**) where spiking is observed with a tunable and countable number of spiking events as shown in Fig. 2(**C**), where for example we have 14 spikes above a certain threshold voltage and the number of spikes can be increased with the addition of some additional bias voltage that alters the coupling strength in order to produce a burst-mode with 4 additional spikes above threshold as shown in Fig. 2(**D**) and thus a total of 18. Fig. S2 in the supplementary as well as Video **S1** show this behavior in real-time as recorded with a digital camera directly on the oscilloscope while performing this experimental measurement. To model the experimental behavior in this burst-mode we advance our non-linear model presented in *(2)* to now consider the case of two or more coupled artificial neurons and introduce a coupling term *(17)* to the differential where $K1_{coupled}v_2(t)$ enters in the equation for $\frac{d^2v_1(t)}{dt^2}$ and $K2_{coupled}\, v_1(t)$ for $\frac{d^2v_2(t)}{dt^2}$:

$$\frac{d^2v_1(t)}{dt^2} + \left(\frac{R_{Mem2}(t)}{R_3} + \frac{C_1}{C_2} + \frac{dR_{Mem2}(t)}{dt}C_1 + \frac{R_{Mem1}(t)}{R_4}\right)\frac{1}{R_{Mem2}(t)C_1}\frac{dv_1(t)}{dt} + \left(\frac{1}{R_3C_2} + \frac{1}{R_3}\frac{dR_{Mem2}(t)}{dt}\right)\frac{1}{R_{Mem2}(t)C_1}v_1(t) + K1_{coupled}\, v_2(t) = 0 \quad (1)$$

$$\frac{d^2v_2(t)}{dt^2} + \left(\frac{R_{Mem4}(t)}{R_5} + \frac{C_5}{C_6} + \frac{dR_{Mem4}(t)}{dt}C_5 + \frac{R_{Mem3}(t)}{R_6}\right)\frac{1}{R_{Mem4}(t)C_5}\frac{dv_2(t)}{dt} + \left(\frac{1}{R_5C_6} + \frac{1}{R_5}\frac{dR_{Mem4}(t)}{dt}\right)\frac{1}{R_{Mem4}(t)C_5}v_2(t) + K2_{coupled}\, v_1(t) = 0 \quad (2)$$

In these coupled non-linear differential equations, the various capacitances $C_x$ and resistances $R_y$ are adjustable elements of the analog circuit to include both intrinsic and extrinsic factors and the $R_{mem}$ elements represent the tunneling memory resistance and any changes observable through the hysteresis. Calculations using this model are shown in Fig. 2(**E**) for the case of two versus three spikes, i.e. the incremental addition of one spike with varying the level of coupling and Fig. 2(**F**) shows the result on a parametric plot of the first vs. second derivatives of the calculated responses and how these burst-modes provide a mechanism for the excitation *(18)* of the qubit states *(19)*.

Next, we interface a carefully calibrated burst-mode neuron network as just described, with a supercooled network of artificial neurons i.e. at 8.2-9.1 degrees Kelvin and that accepts the burst-mode spiking as the exciting source. Fig. 3(**A, B**) shows measured time-domain spectra test results for near 40-70 milli-second durations as well as data acquired in the associated phase-plane that represents the degrees of freedom of each neuron qubit with the spiking sequence produced with the lead artificial neuron and optimized as shown with the sequence shown with milli-second scale widths. Fig. S(**2-4**) in supplementary Video **S2** shows further raw oscilloscope captured data taken as screenshots as well as in video format during our live experimentation in the laboratory. Fig. 3(**C, D**) shows a further example, where we isolate an interesting case now that shows in time-delay of 14.2 milli-seconds in the behavior as it pertains to the two degrees of freedom and points to the impacts of non-Markovianity in the system. To model the behavior reported here under these experimental conditions, we advance our model to consider an angular dependence, therefore calculating using a Hamiltonian function we therefore upgraded with a further rotational dependence $\theta_k$ that represents experimentally a unique pulse sequencing.

$$H_{neuron_{qbit}}(t,\theta) = \sum_k^N \left( -gsin\left(\frac{\theta_k}{2}\right)\left(\sigma_{ge}^\dagger a + a^\dagger \sigma_{ge}\right) - Acos\left(\frac{\theta_k}{2}\right)\left(\sigma_{ge}^\dagger + \sigma_{ge}\right)\sin\left(\left(\frac{t}{e}\right)^2\right)\right) \quad (3)$$

The observed qubit-states across the Bloch sphere- when considering open quantum system dynamics *(20)* with the master equation approach and a longer-term non-Markovianity treatable with transfer tensor approaches *(21)*- that effectively approximates a long duration of learning time *(22)*. We perform extensive signal processing analysis of the measurement data as shown in Fig.

3(**E**) and calibration of this quantum model to provide a good agreement between the reconstructed output and the observed behavior and necessary to consider a degree of non-Markovianity in modeling the quantum memory learning process.

## Process to direct information generation under memory non-Markovianity

As an application of these neural-networks, and with the calibrated models developed from experimental measurements, we apply this to the concept of the generating and routing process of information packets. We examine the scenario under memory non-Markovianity as that provides novel capability *(23-25)*. We analyze the case of a coupled and entangled set of these synthetic artificial neuron qubits and we introduce time-dependent coupling terms *(26-29)* for the neuron qubit states and excitation sources $q_1$ and $q_2$: $\sigma_{ge_q1}, \sigma_{ge_q2}, \sigma^{\dagger}_{ge_q1}, \sigma^{\dagger}_{ge_q2}, a^{\dagger}_{q1}, a_{q2}$. Fig. 4(**A, B**) shows the Bloch sphere representation of these coupled and entangled states and the extracted coherence from the first order correlation function in Fig. 4(**C**). In accordance with our measurements in Fig. 3(**C**) that show signs of non-Markovianity, our calculations in Fig. 4(**D**) show these projections including for the degrees of freedom expectation for the case on a longer-term memory and the visualization on the Bloch sphere (30). With these measurements and simulated extrapolations/extractions we define in Table I the algorithmic dependencies of the qubit-neuron projected degrees of freedom in relation to the level of coherence and whether they are in weak or strong entanglement and how that affects the system's cognitive-decision significance for generating regular information packets as well information attributes that represent enhanced awareness and a full awareness suitable for subsequent routing or use of the quantum process *(31)* for assisting the generation of network packet parameters and a matrix of the form shown here is used to determine the decision-making aspects of the information packet's routing that's calculated from the integrated density matrices and to extract the level of non-Markovianity.

$$\begin{pmatrix} \rho_{11} & \rho_{12} \\ \rho_{21} & \rho_{22} \end{pmatrix} = \frac{1}{3}\begin{pmatrix} \rho_{11_q1} & \rho_{12_qbit1} \\ \rho_{21_qbit1} & \rho_{22_qbit1} \end{pmatrix} + \frac{1}{2}\begin{pmatrix} \rho_{11_qbit2} & \rho_{12_qbit2} \\ \rho_{21_qbit2} & \rho_{22_qbit2} \end{pmatrix} + \frac{1}{6}\begin{pmatrix} \rho_{11_qbit3} & \rho_{12_qbit3} \\ \rho_{21_qbit3} & \rho_{22_qbit3} \end{pmatrix} \quad (4)$$

Fig. 5(**A-C**) shows an example of this algorithm implemented for the situation of generating three types of information packets and Fig. 5(**D-F**) shows an integrated process where in real time the decision-making ability impacted from the level of awareness or non-Markovianity determines the dynamical behavior from the $\theta_{k1 \to k3}$. Videos of this experiment are presented in Videos (**S3-S4)**.

## Conclusion:

We examined the built-in and reconfigurable quantum states of qubit-neurons built to operate specifically with co-integrated burst-spiking neurons. The system represents a biological-inspired network of neurons where quantum information algorithms are performable in conjunction with traditional spiking encoding. The artificial neuron types used here comprise networks of superconductor-ionic memories that provided sufficient coherence time for the quantum memory properties to impact the quantum trajectories We used direct measurements and reconstructed behavior to calibrate a quantum model to examine the use of a set of entangled qubit neurons to enable the generation of novel information packets and to use the measures of their coherence/entanglement as well as non-Markovianity to define information awareness parameters. This work is thus of crucial importance to the development of quantum neuromorphic hardware.

**Supplementary Materials:**

**The PDF file includes:**

Materials and Methods

Supplementary Figures

Figs. S1-S3

**Auxiliary Supplementary Materials for this manuscript include the following:**

Videos S1 to S4

**Acknowledgments:**

We're grateful for the NIWC Pacific information technology and laboratory infrastructure.

**Funding:** This work was funded by Naval Information Warfare Center (NIWC) Pacific. Distribution Statement A. Approved for public release: distribution is unlimited.

**Author contributions:** Conceptualization: OMN, Methodology: OMN, Investigation: OMN, CSH, Validation: OMN, Formal Analysis: OMN, Resources: OMN, Data Curation: OMN, Visualization: OMN, Writing – original draft: OMN, Writing – review & editing: OMN, CH

**Competing interests:** OMN declares US Patents (1) "Reconfigurable, tunable quantum qubit circuits with internal, nonvolatile memory", US Patent 9,755,133 B1 and (2) "Quantum computer based method for analyzing cyber data and spectra while performing optimization based on the analysis", US Patent US11,373,112 B2. CH doesn't have competing interests.

**Data and materials availability:** All data are available in the main text or supplementary materials." Several videos are included in the auxiliary supplementary materials. Requests for additional data or questions should be made to the corresponding author.

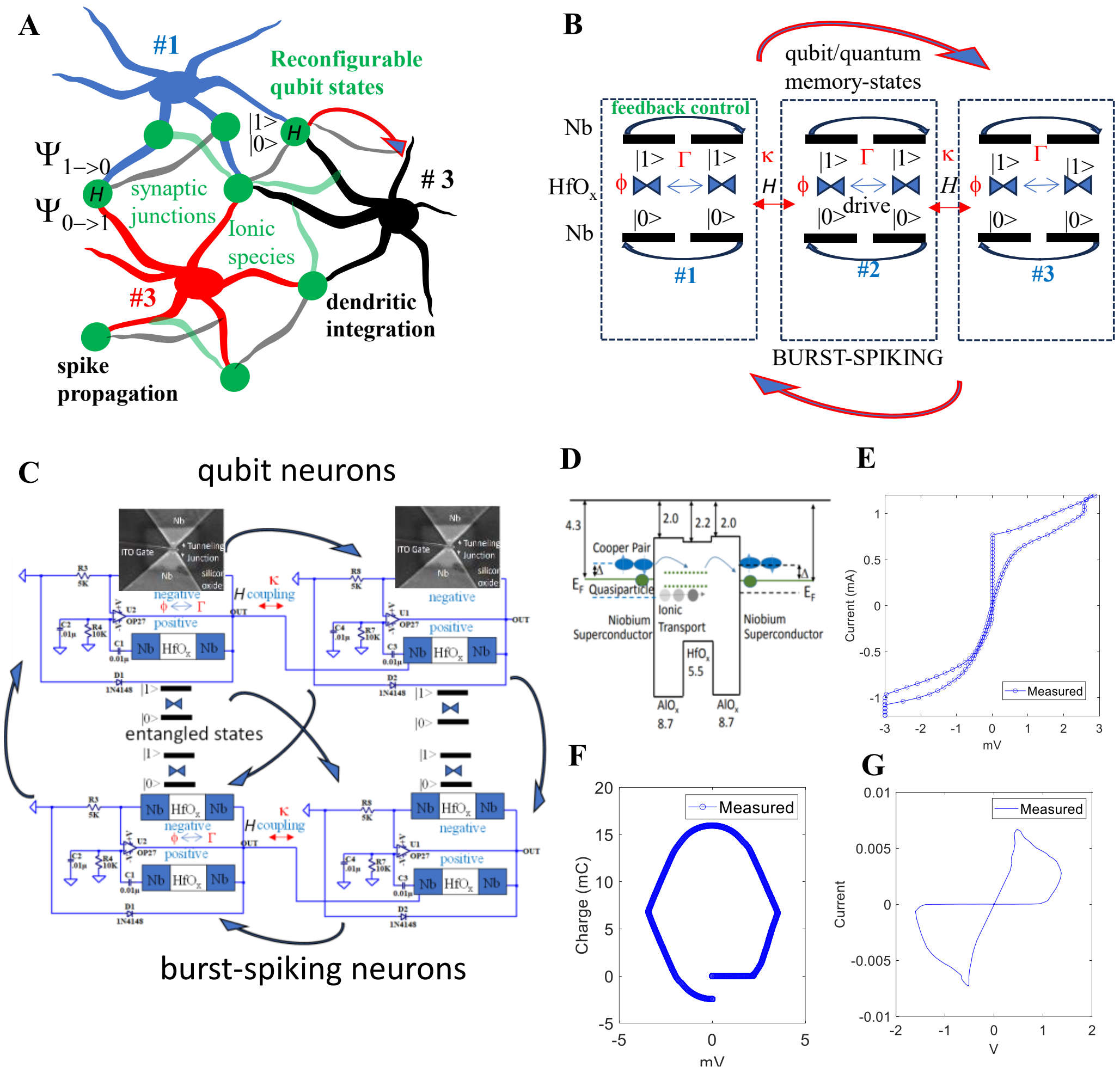

**Fig. 1:** (A) Biological-inspired illustration of the synthetic artificial neuron networks coupled and entangled qubit states that are inputted with the co-integrated burst-mode spiking representing the ionic synaptic junctions and dendritic integration that provide for the inter-neuron coupling and feedback control process. (B) Energy-level description of the qubit/quantum memory states that are available for performing quantum information algorithms and the respective energies and intercoupling strengths that are connected to the circuit parameters (C) Circuit schematic (modified from ***(1)*** that integrated two qubit neurons and two burst-spiking neurons and where variations can be made to produce a desired response. (D) Energy-bands for the Cooper pairs/quasiparticles (reproduced with permission from ***(10)***. *(*E) Superconducting quantum tunneling conductance and (F) ferroelectric hysteresis measurement (G) Memristive response of a co-integrated burst-spiking artificial neuron.

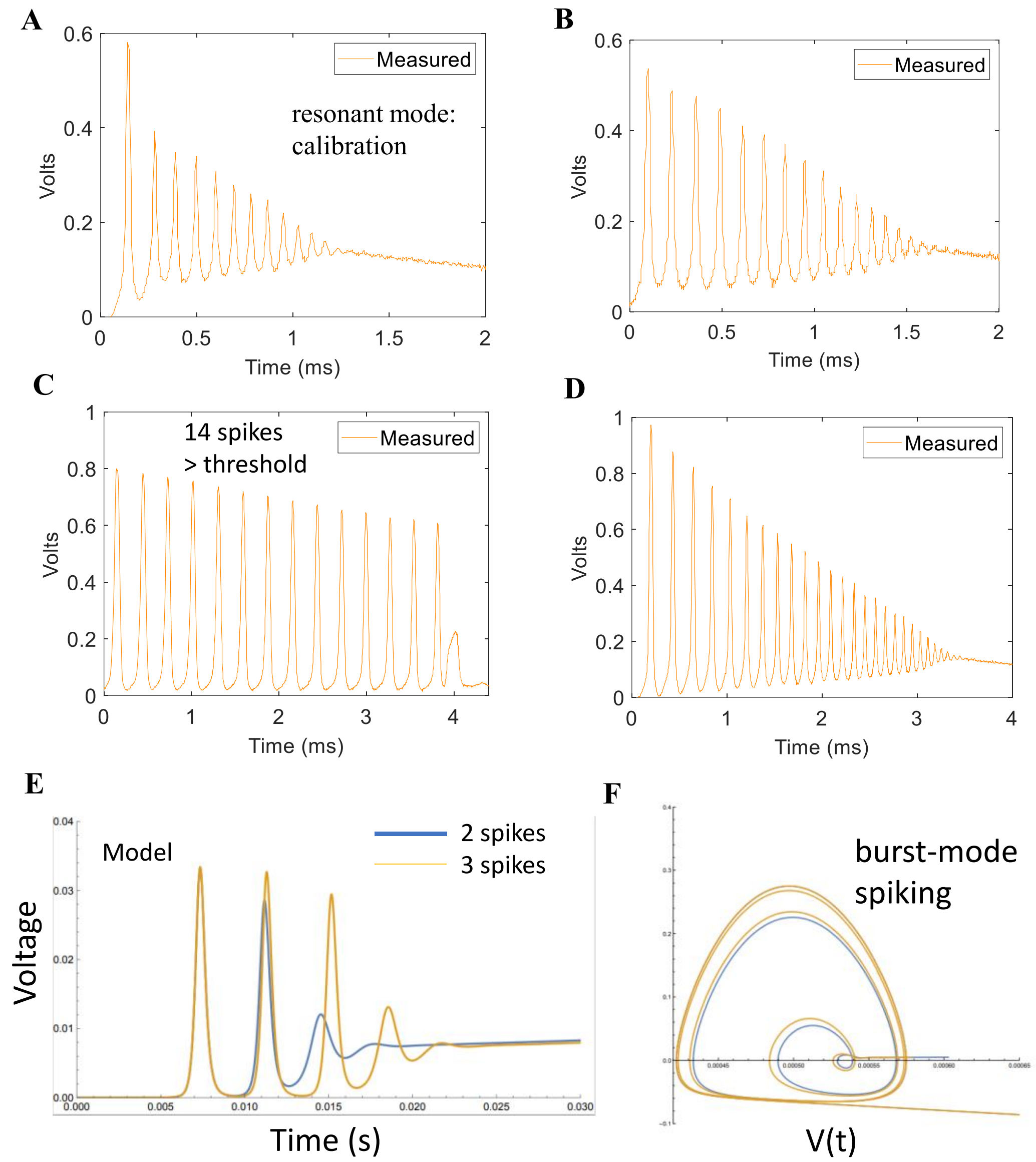

**Fig. 2**: Measurements of the synthetic artificial neuron network in burst-spiking mode (A-B). Calibrations with 8-10 spike counts and at room temperature with example showing how varying the neuron-neuron coupling strength results in incremental changes to the number of spikes from (C) 14 above a threshold and (D) an increase to 15 and 16 spikes above threshold thus demonstrating the ability to control the excitation of the co-integrated neuron-qubits in this synthetic biological inspired system (E) Modeled results with coupled non-linear differential equations 3 and 4 using our updated mathematical model that integrates an adjustable inter-neuron coupling coefficient varying the spike count incrementally from 2 to 3 in this case that is computationally feasible and (F) Parametric plot of the first and second derivatives the spiking output with the commensurate incremental increase in the spikes in such a highly non-linear regime.

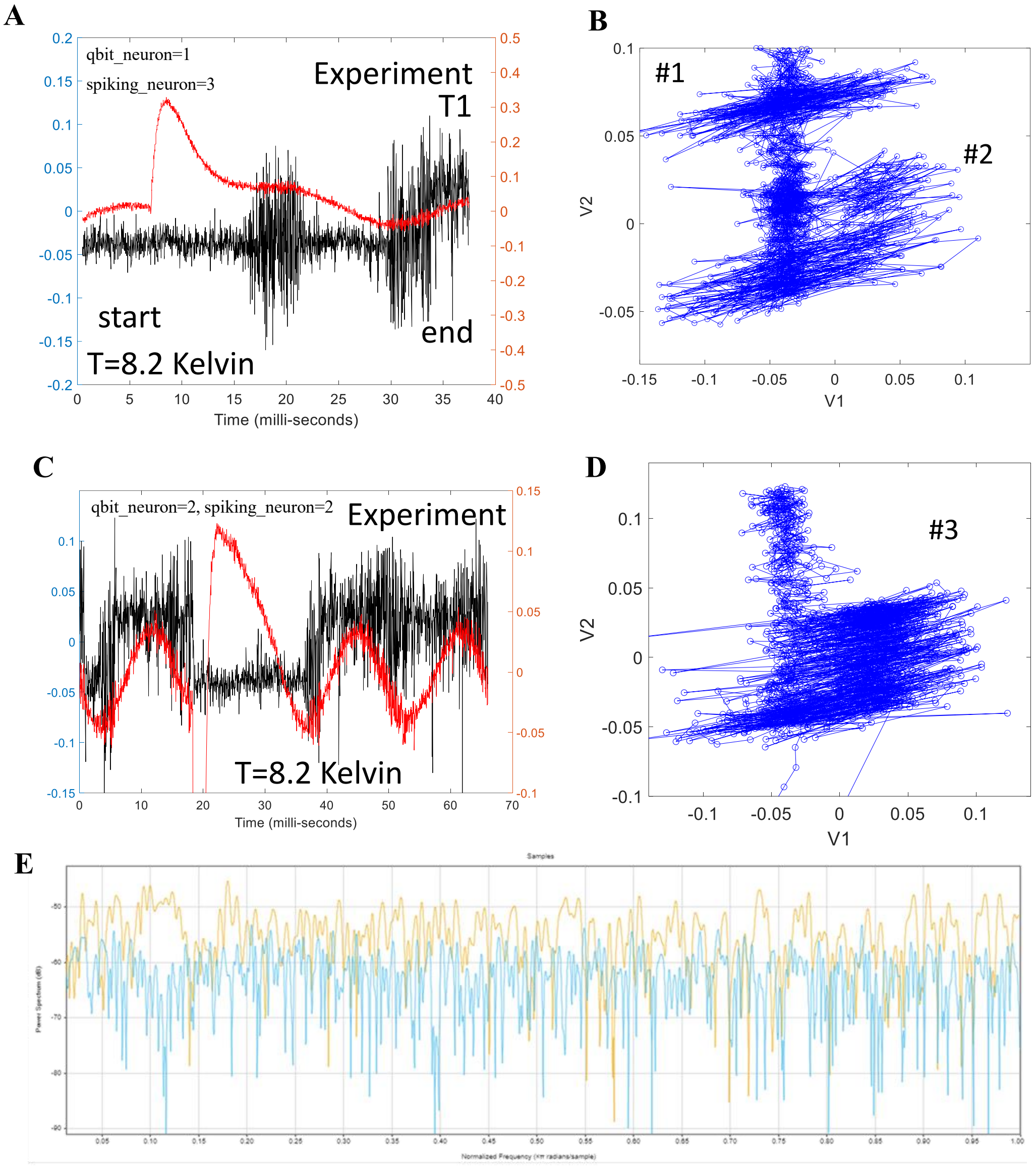

**Fig 3:** (A) Experimental measurements of the time-domain response of the qubit neurons while cooled to 8.2 Kelvin in the superconducting regime and measured at the appropriate nodes that represent the qubit-states trajectories during sustainment of a coherent processes while under the unique pulsed excitation sequences that apply the burst mode dynamics. T1 extracted from logarithmic analysis of the decay function is 7.4 milli-seconds and enhanced due to the feedback coupled environment and where the signatures of the information packets#1 and #2 are annotated (B) Measurements done in the phase plane for the same situation and (C) a different case where the burst mode dynamics are further intensified and where the number of burst-spiking neurons and qubit-neurons are both equal to two showing information packet #3 (D) Phase plane measurements and in (E) Frequency domain plots of the output that are used for reconstructing the quantum states.

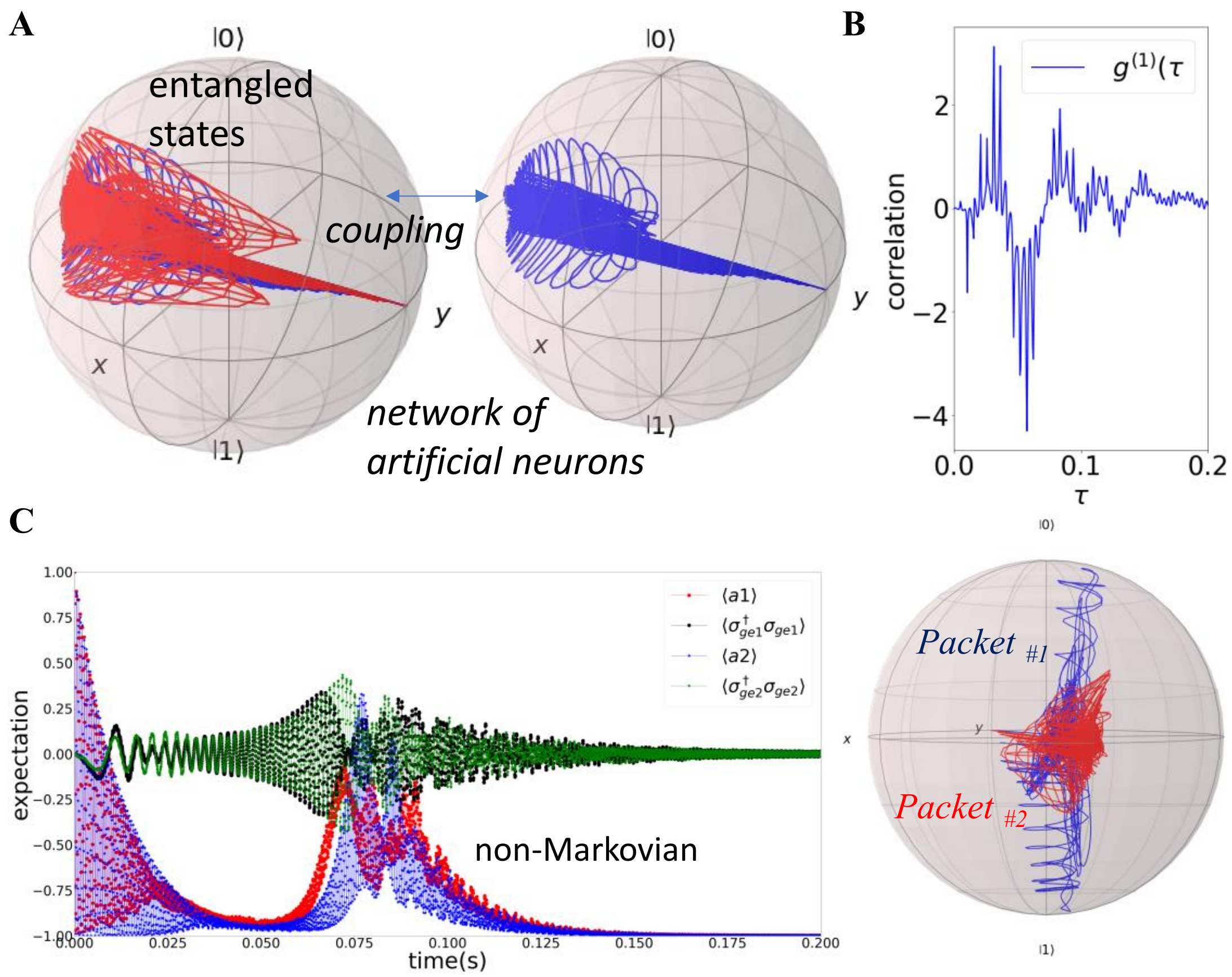

**Fig 4:** (A) Modeled and reconstructed Bloch-sphere visualization of the qubit state trajectories of a coupled and entangled set of qubit neurons with changing burst-mode spiking durations i.e. number of spike counts using our quantum model with an adjustable angle dependence in the Hamiltonian (B) Extracted level of first-degree quantum correlation during the dynamical event with a coherence time showing sustainment over the course of duration (C) Projection of the entangled states degrees of freedom as a function of time for the case of non-Markovianity observed with the signs of the information flow due to sustained quantum memory coherence of the qubit-states from the co-integration of the superconductor-ionic memories that are integrated in a feedback coherence protection environment and where the burst-spiking excitation provides for further shielding of the quantum states during the dynamical read-out and the Bloch-sphere visualization of the quantum information states trajectories and the resulting information packet generation and routing trajectory behavior that represent the experimentally measured packets#1 and #2.

| Table I | | |
|---|---|---|
| qubit-neuron network's correlation state | cognitive-decision significance | non-Markovianity degree of information (normalized) |
| A) 3< $q_1$<8 (strong entanglement) | • Regular packet -> continue generation | • level > 1.23, with a correlation>85% |
| B) 14< $q_2$<20 (medium entanglement) | • Enhanced awareness -> probe & read-out | • level > 0.21 with a correlation>65% |
| C) 22< $q_3$<23 (low entanglement) | • Elevated-> full route & reset | • level >0.052 with a correlation>55% |

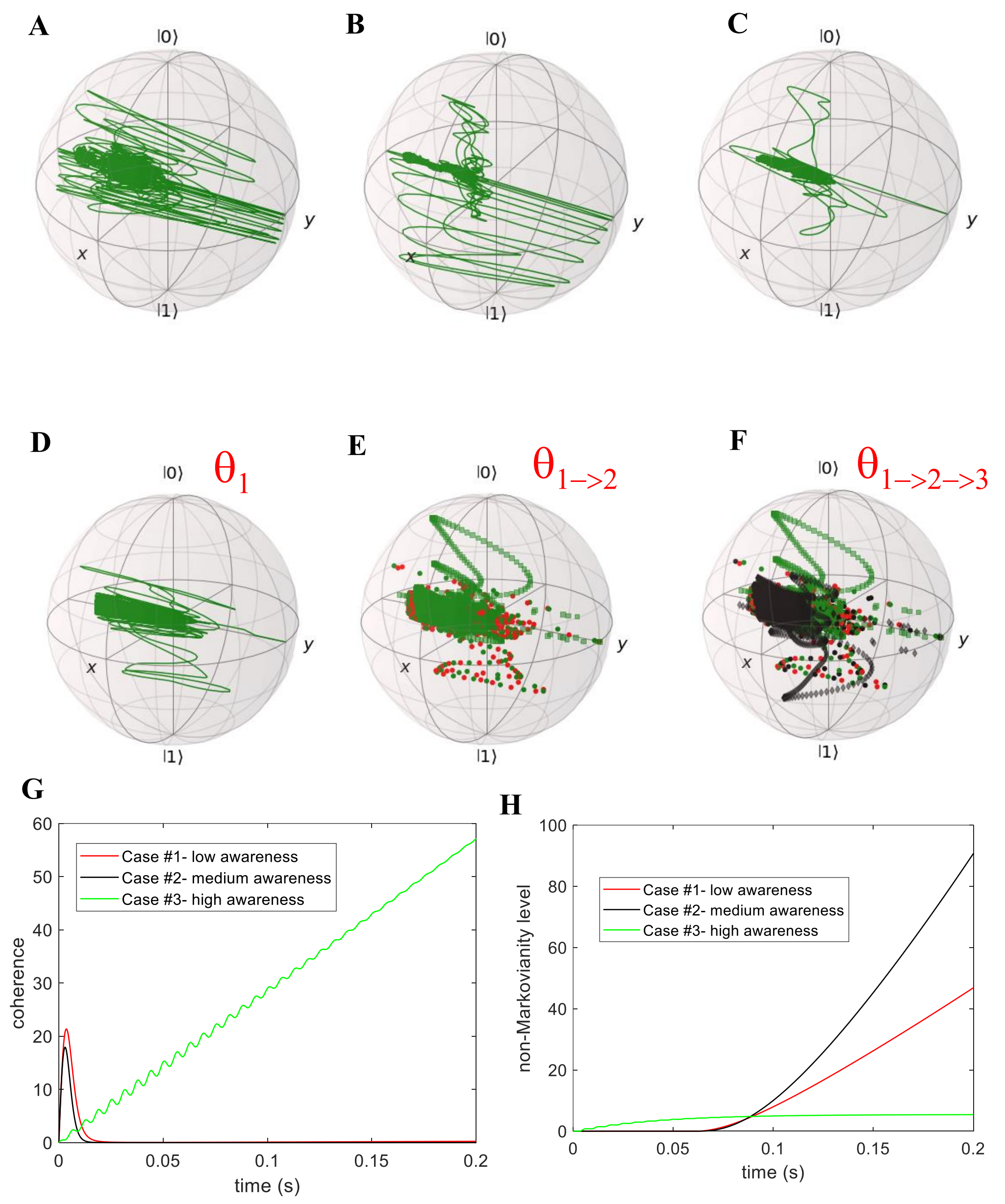

**Fig. 5**: (A) Bloch sphere representation of reconstructed qubit states and quantum trajectories with this process for information generation taking a simple example of three scenarios that represent low/medium/high levels of awareness and with a non-Markovian model used in accordance with the directives of Table I. The plots in (A-C) show the trajectories and real-time changes in the temporal trajectories as a function of adjustments due to the level or degree of non-Markovianity as well as the level of sustained correlation or coherence time of sets of entangled qubit-neurons where a quantum process directs this generation and routing thus effectively providing a sense of alertness to the information packets as determined from the inter-neuron-qubit coupling coefficients by the burst-mode spiking outputs and angle dependence thereby properly contributing to the overall networks ability to generate as well as learn more novel types of information as shown for examples (D-F) the extracted level of (G-H) correlation and non-Markovianity.

# Supplementary Materials for

## Reconfigurable qubit states and quantum trajectories in a synthetic artificial neuron network with a process to direct information generation from co-integrated burst-mode spiking under non-Markovianity

**Authors:** Osama M. Nayfeh[1†], Chris S. Horne[1]

**Affiliations:** [1] Naval Information Warfare Center (NIWC) Pacific, San Diego, CA USA

**†Corresponding Author:** osama.m.nayfeh.civ@us.navy.mil

**The PDF file includes:**

Figs. S1-S3

**Auxiliary Supplementary Materials for this manuscript include the following:**

Videos S1 to S4

**Materials and Methods**

**Supplementary Figures**

## Materials and Methods

The synthetic quantum neurons integrated superconductor-ionic quantum memories. The circuits examined here integrated coupling elements that produce co-integrated burst spiking modes, and the degree of coupling correlates the level of entanglement between the qubit neurons and the qubit states are reconfigurable and their quantum trajectories can be monitored across the Bloch sphere by means of co-integrated burst-spiking neurons. Depending on the levels of resulting entanglement and the degree of coherence the information generated can produce so called regular information packets are those with heightened awareness depending on the quantum process and the configuration of the quantum states and the extracted non-Markovianity.

## Supplementary Figures

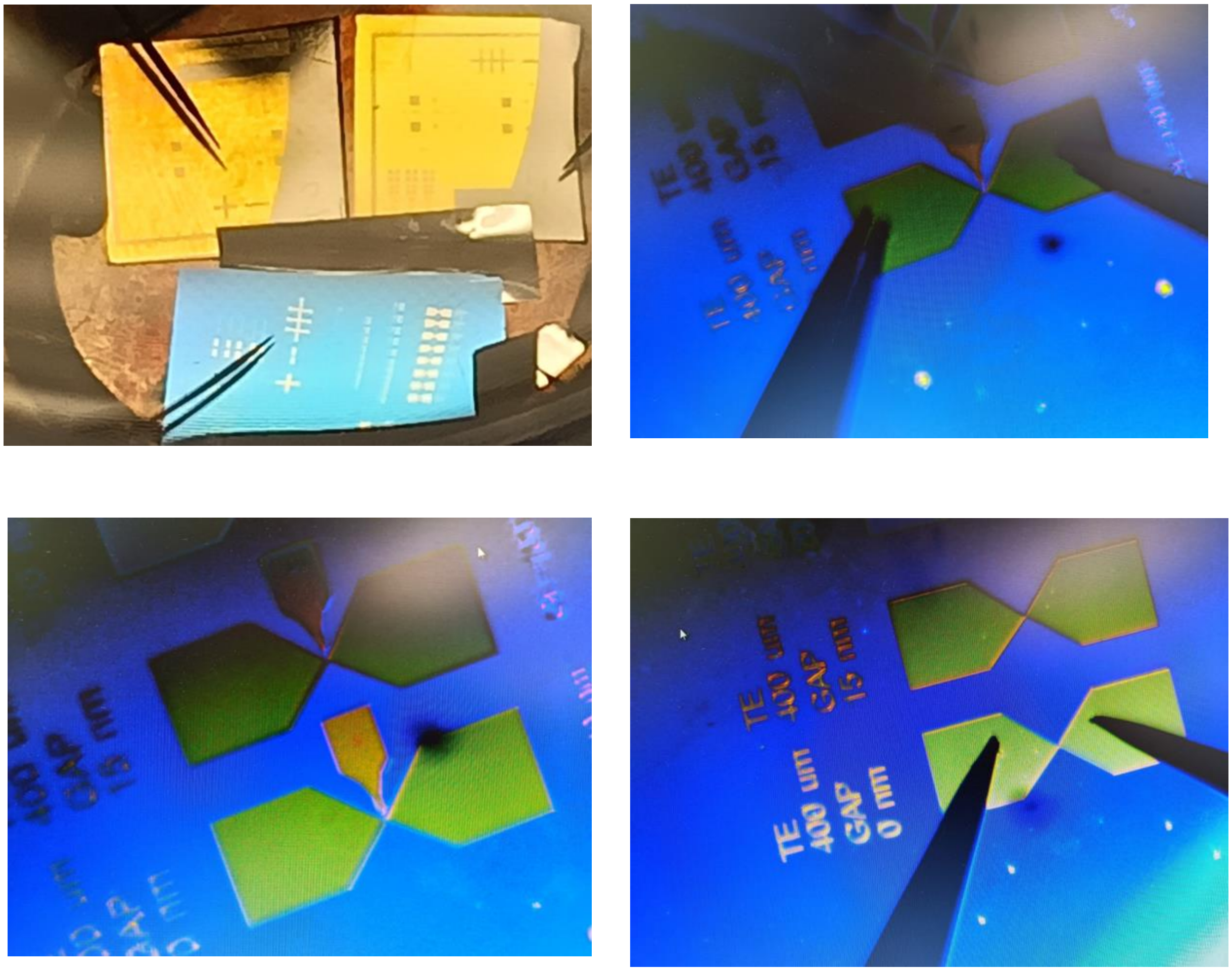

**Fig S1**. Micrographs of qubit-neuron and burst-spiking neuron chips while electrically probed during their operation in these circuits.

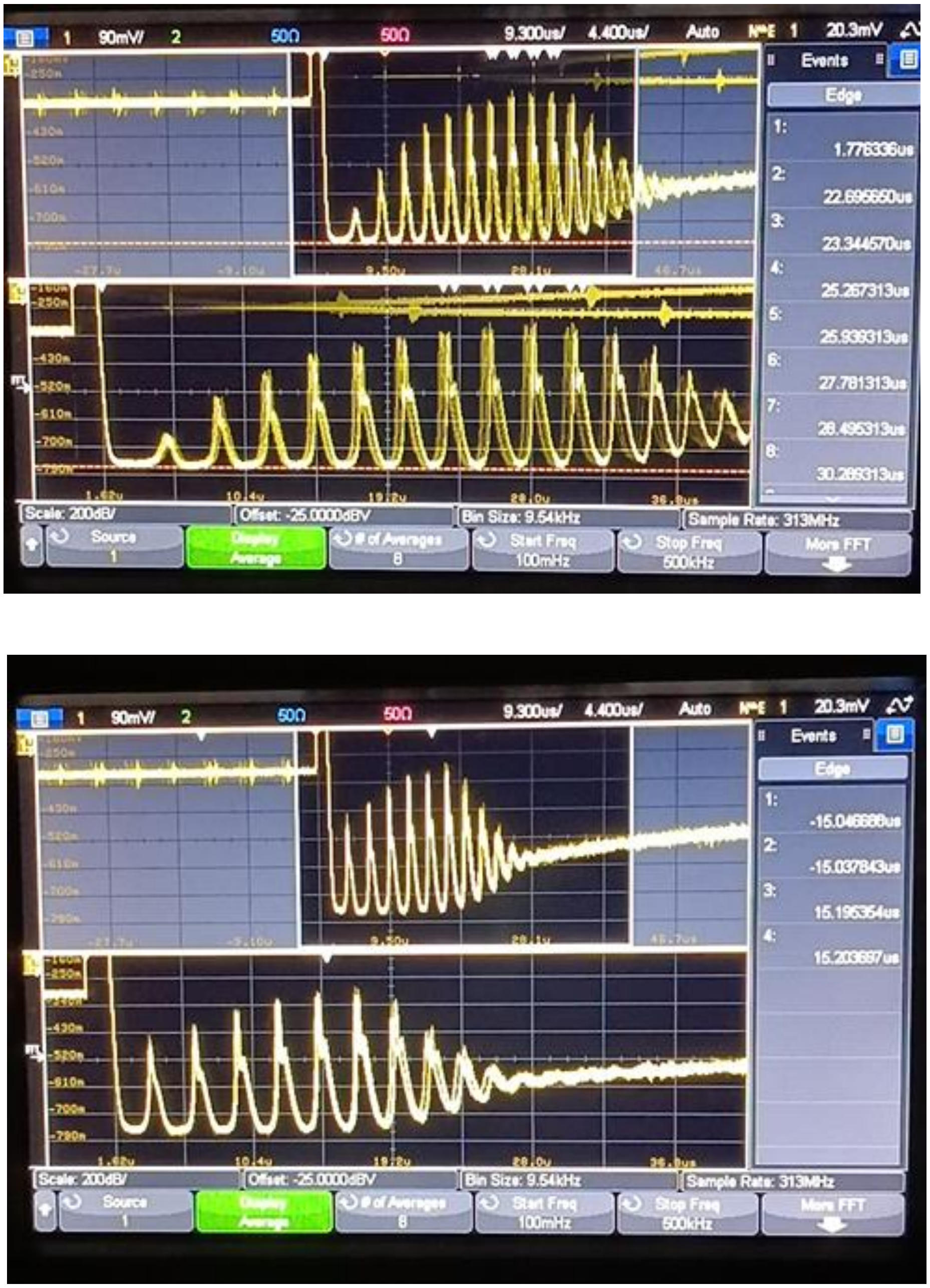

**Fig S2. (A)** Raw experimental measurement data captured while observing the co-integrated spiking modes and by adjusting the level of coupling between the synthetic neurons the capability to count the addition or subtraction of the number of spikes from more to less depending on the matching level of the specific desired qubit state. Also see Video S2 for live captured images of these dynamics. **(B)** Observance of the accompanying software system that shows an example of how an information packet can be generated in this case and directed by the quantum processes examined here and suitable for propagation on a conventional interface.

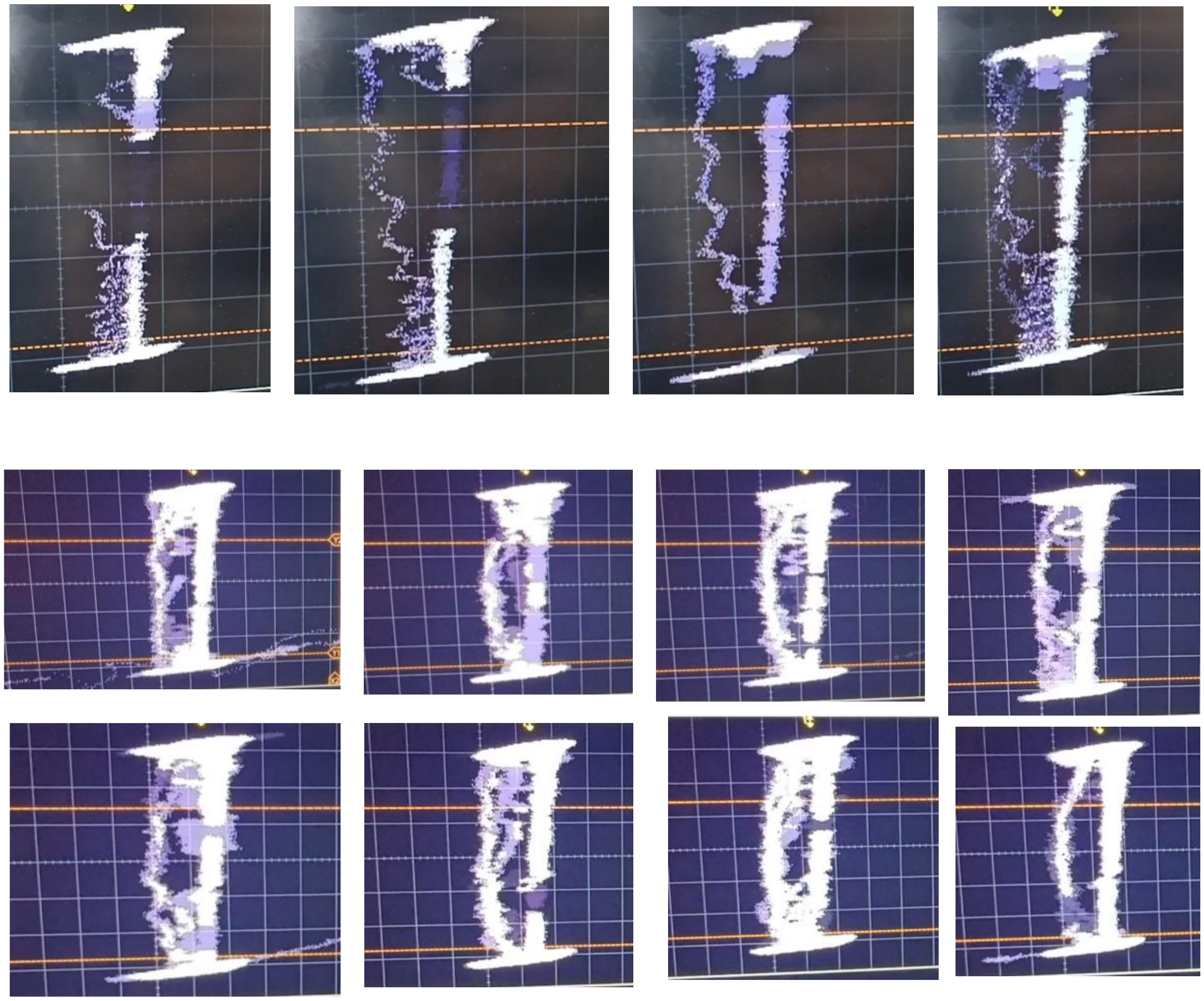

**Fig S3.** Raw experimental measurement data captured in the phase plane while reconfiguring the quantum trajectories of the qubit states for a couple of cases of information generation **(A)** For a time duration of 500 nanoseconds and **(B)** For a time duration of 1300 microseconds. The complex trajectories are observed and are dependent on the level of entanglement between the qubit neurons and their coherence times as determined by the quantum memory quality. The impacts of non-Markovianity are apparent. Also see Video S1 for live captures images of these dynamics.